\documentclass[conference]{IEEEtran}
\usepackage{cite}
\usepackage{amsmath,amssymb,amsfonts}
\usepackage{algorithmic}
\usepackage{graphicx}
\usepackage{textcomp}
\usepackage{xcolor}
\usepackage[hyphens]{url}

\usepackage{xspace}
\usepackage{fancyhdr}
\usepackage{caption}
\usepackage{subcaption}
\usepackage{csquotes}
\usepackage{verbatim}
\usepackage{paralist}
\usepackage{subfiles}
\usepackage{enumitem}
\usepackage{soul}
\usepackage{booktabs}
\usepackage{tabularray}
\usepackage{color}
\usepackage{multirow}
\usepackage{multicol}

\def\BibTeX{{\rm B\kern-.05em{\sc i\kern-.025em b}\kern-.08em
    T\kern-.1667em\lower.7ex\hbox{E}\kern-.125emX}}

\newcommand\eat[1]{}
\newcommand{\acceleratorname}{PhotoHDC\xspace}
\newcommand{\acceleratornametrain}{PhotoHDC\xspace}

\DeclareMathOperator*{\argmax}{\arg\!\max}
\newcommand{\rxc}{$R\times C$\xspace}
\newcommand{\firstEnc}{traditional\xspace}

\title{Towards Efficient Hyperdimensional Computing Using Photonics \vspace{-0.2in}}

\author{
\IEEEauthorblockN{Farbin Fayza\IEEEauthorrefmark{1}, 
Cansu Demirkiran\IEEEauthorrefmark{1},
Hanning Chen\IEEEauthorrefmark{2},
Che-Kai Liu\IEEEauthorrefmark{3},
Avi Mohan\IEEEauthorrefmark{1}}
\IEEEauthorblockN{
Hamza Errahmouni\IEEEauthorrefmark{2},
Sanggeon Yun\IEEEauthorrefmark{2},
Mohsen Imani\IEEEauthorrefmark{2},
David Zhang\IEEEauthorrefmark{4},
Darius Bunandar\IEEEauthorrefmark{5},
Ajay Joshi\IEEEauthorrefmark{1}\IEEEauthorrefmark{5}
}
\IEEEauthorblockA{\IEEEauthorrefmark{1}Boston University}
\IEEEauthorblockA{\IEEEauthorrefmark{2}University of California, Irvine}
\IEEEauthorblockA{\IEEEauthorrefmark{3}Georgia Institute of Technology}
\IEEEauthorblockA{\IEEEauthorrefmark{4}SRI International}
\IEEEauthorblockA{\IEEEauthorrefmark{5}Lightmatter}
}

\begin{document}
\maketitle
\thispagestyle{plain}
\pagestyle{plain}


\begin{abstract}
Over the past few years, photonics-based computing has emerged as a promising alternative to CMOS-based computing for Machine Learning (ML)-based applications, in particular, for Deep Neural Networks (DNNs).
Unfortunately, the non-linear operations and the high-precision requirements of DNNs make it challenging to design efficient photonics-based computing systems for DNNs.
Hyperdimensional Computing (HDC) is an emerging, brain-inspired ML technique that enjoys several advantages over DNNs, including (i) being lightweight, i.e., involving simple linear algebraic operations; (ii) requiring low-precision operands; and (iii) being robust to noise introduced by the nonidealities in the hardware.

In this paper, we argue that photonic computing and HDC complement each other effectively.
We propose \acceleratorname, the first electro-photonic accelerator for HDC training and inference, supporting traditional, record-based, and graph encoding schemes.
Our novel accelerator microarchitecture utilizes Mach-Zehnder Modulators and photodiodes to accelerate HDC encoding, bundling, and similarity measurement operations while minimizing the use of digital modules for other system components.
Our analyses show that \acceleratorname can achieve two to five orders of magnitude lower Energy-Delay Product (EDP) than the state-of-the-art electro-photonic DNN accelerators when running HDC training and inference across popular datasets.
We also show that photonics can overcome the challenges faced by Compute-in-Memory-based HDC accelerators and achieve multiple orders of magnitude lower EDP for both HDC training and inference.

\end{abstract}

\section{Introduction}\label{sec:intro}

Photonics-based computing presents a compelling alternative to CMOS-based computing for Machine Learning (ML) applications, especially as Moore's law reaches its limits~\cite{sunny2021survey}. 
Photonic chips offer significantly higher throughput and energy efficiency than their electronic counterparts, particularly for performing Matrix-Vector Multiplication (MVM)~\cite{zhou2022photonicgemm}---which is a key operation in Deep Neural Networks (DNNs).
Many recent studies have showcased the potential of photonics in accelerating DNN inference, yielding remarkable results~\cite{zhou2022photonicgemm}\cite {shiflett2021albireo}\cite{shiflett2020pixel}\cite{zhu2024lightening}\cite{cansu2021electro}\cite{sunny2021crosslight} \cite{liu2019holylight}.
Although these photonics-based designs are promising for DNN acceleration, there are a few limitations to this approach.  
First, DNNs contain a variety of non-linear operations (ReLU, GeLU, etc.) that cannot be efficiently performed in photonics and are typically executed using electronic circuits.
While the majority of DNN operations are linear, these non-linear operations can easily become the bottleneck in the system as electronic circuits struggle to match the high bandwidth of photonic circuits. 
Second, the analog noise inherent in photonic circuits, the process variations, and the fabrication errors limit the achievable precision in photonic circuits. 
High-precision data converters also come with substantial power costs and are impractical to use in energy-efficient analog systems. 
Using low-precision fixed-point arithmetic ($\leq8$-bits), however, can cause a significant drop in the DNN accuracy, especially in today's large DNNs~\cite{demirkiran2023blueprint}.
Essentially, to maximize the benefits of photonics for AI, an ideal computing model should use minimal non-linear operations and be able to achieve high accuracy using low-precision operands.


Hyperdimensional Computing (HDC) has recently emerged as a brain-inspired ML technique whose accuracy is comparable to DNNs in various applications while requiring less computational resources~\cite{ge2020hdc-lassification-review}.
HDC maps sensory input data into high-dimensional hypervectors (${\approx} 10^4{-}dimensions$) and relies on simple linear operations such as multiplication and addition between these hypervectors~\cite{rosing21theoryHDC}.
Unlike traditional DNNs with deep and complex networks trained via back-propagation, HDC employs simpler architectures, thereby enhancing computational efficiency and interpretability \cite{imani21scalableHDC}.
In addition, HDC can operate with much lower precision than DNNs while still achieving comparable accuracy~\cite{rahimi2017high}\cite{ge2020hdc-lassification-review}.
Previous research demonstrates HDC's applicability across various tasks such as image classification, speech recognition, object tracking, face detection, and robotic tasks~\cite{ge2020hdc-lassification-review, imani2022neural, neubert2019introduction}.
Moreover, in contrast to DNNs, HDC is extremely robust to noise and can retain acceptable accuracy levels across a wide range of hardware failures~\cite{rahimi2017high, kazemi2021mimhd}.

\textbf{In this paper, we argue that HDC's remarkable tolerance for low-precision operations and reliance on linear computations align perfectly with the inherent strengths of photonics.}
We believe that a combined use of these technologies holds immense potential for future computing advancements.
Currently, Non-Volatile Memory (NVM)-based Compute-in-Memory (CiM) is recognized as the most efficient platform for HDC~\cite{karunaratne2020memory, zou2022biohd, baghdadi2021dual, kazemi2021mimhd, liu2022cosime, kazemi2022achieving}.
CiM successfully eliminates the data movement cost and achieves better performance compared to traditional Von Neumann architecture-based hardware accelerators, such as ASICs and FPGAs~\cite{barkam2023hdgim}.
However, it poses several challenges that photonics can address.
First, CiM is not efficient in handling multi-bit operations, which limits the achievable accuracy in HDC for complex tasks.
Second, updating the values in an NVM array is costly and time-consuming ($\approx$ tens to hundreds of ns)~\cite{xu2015overcoming, yu2021rram}. 
While, for inference, this cost can be amortized by loading all the required values into the NVM array upfront and reusing them for multiple queries, these values need to be updated frequently for training, which causes long delays between operations. 
Lastly, the low endurance of NVM cells limits the number of writes in the NVM array~\cite{yu2021compute}. 
Therefore, while CiM can be used in edge devices for executing inference on a pre-trained and fixed model,
for data center/cloud-scale systems that perform both training and inference across various applications, CiM is not a feasible option.
Unlike NVM cells, photonic devices can achieve $\geq8$-bit precision during operations~\cite{cansu2021electro}\cite{zhu2024lightening}, can attain high-bandwidth modulation with rates reaching tens of GHz\cite{MZM_Akiyama:12,PD_sheng2010ingaas}, and do not suffer from endurance issues~\cite{sun2022silicon,seok2016large}. 
Current electro-photonic DNN accelerators can \emph{technically} run most HDC models as many of the HDC operations can be represented as MVM operations. 
However, using such accelerators as is for HDC models results in poor efficiency and low utilization as these accelerators are specialized for DNN workloads. 
In particular, for HDC encoding schemes requiring $O(N^2)$ data modulation per cycle for an $N\times N$ optical array (e.g., record-based or graph encoding), current Micro-Ring Resonator (MRR)- and Mach-Zehnder
Interferometer (MZI)-based approaches result in impractical modulation costs. 
Moreover, photonic tensor cores relying on Singular Value Decomposition (SVD)~\cite{linearalgenbrastrang1993}\cite{cansu2021electro} struggle with iterative HDC training as SVD, which is a costly operation, needs to be performed for each iteration. 
To be able to support the wide range of data-intensive operations in HDC, a general HDC accelerator must offer (1) low-cost data modulation at high frequency and (2) flexibility for versatile HDC operations involved in training and inference with various encoding schemes.

In this paper, we propose---for the first time---a novel electro-photonic accelerator, \acceleratorname, for HDC training and inference.
The microarchitecture of \acceleratorname includes an array of multiply-accumulate (MAC) units, each based on a Mach-Zehnder Modulator (MZM) and a Photo-Detector (PD) pair, low-precision data converters, memory units, and minimal additional digital circuitry.
The MAC units consisting of MZMs and PDs enable high-speed modulation with minimal cost to support efficient HDC operations.
In addition, we provide flexibility in the dataflow for optimizing the performance of various HDC operations.
Our design offers different routing options specialized for each HDC operation to maximize the number of high-speed operations in the analog domain and minimize the number of analog-to-digital conversions.
The key contributions of our work are as follows:
\begin{itemize}[leftmargin=*]
    \item We, for the first time, propose using photonics as a computing platform for HDC applications. We argue that photonics and HDC make a good match as HDC mainly involves linear operations that can be performed efficiently in photonics and it is extremely resilient to low-precision operations.
    \item We propose \acceleratorname, the first ever electro-photonic accelerator
    for HDC training and inference. 
    Addressing the various demands of HDC operations, we design our all-in-one accelerator to support three distinct HDC encoding algorithms: traditional, record-based, and graph encoding, and other HDC operations such as bundling, and similarity measurement, all within the same accelerator using different datapaths.
    
    \item We introduce novel dataflows, designed to work with each encoding scheme, for both HDC training and inference on \acceleratorname.
\end{itemize}

We evaluate the performance of \acceleratornametrain across frequently employed datasets for HDC models.
We also map the HDC models to three popular State-Of-The-Art (SOTA) electro-photonic accelerators for DNNs: ADEPT~\cite{cansu2021electro}, ALBIREO~\cite{shiflett2021albireo}, and DEAP-CNN~\cite{deap-cnn}, and provide a comprehensive comparison highlighting the limitations of these accelerators, when running HDC workloads.
For both HDC training and inference, \acceleratorname achieves two to five orders of magnitude lower Energy-Delay Product (EDP) while being one order of magnitude more area efficient on average than SOTA electro-photonic DNN accelerators.
We also compare \acceleratorname against CiM-based accelerators. 
Our analyses show that \acceleratorname can achieve four orders of magnitude lower EDP than CiM-based accelerators for both HDC training and inference.

\section{Background}\label{sec:background}
\subsection{HDC Overview}\label{subsec:background_hdc}




Similar to DNNs, HDC consists of training and inference steps.
Both HDC training and inference require an encoding step where the input data is mapped to a hyperdimensional space. 
In this paper, we investigate three popular encoding schemes:~\enquote{traditional} encoding~\cite{hernandez2021onlinehd,kim-date-2023efficient,khaleghi2021tiny}, \enquote{record-based} encoding~\cite{imani2017voicehd,kazemi2021mimhd,imani2018hierarchical} and \enquote{graph} encoding~\cite{poduval2022graphd,nunes2022graphhd}. 
Below we first describe the training and inference processes in detail using \firstEnc encoding, and later explain record-based encoding and graph encoding and highlight how they differ from \firstEnc encoding.

\subsubsection{HDC Training}
\begin{figure}[t]
    \centering
     \includegraphics[width=0.48\textwidth]{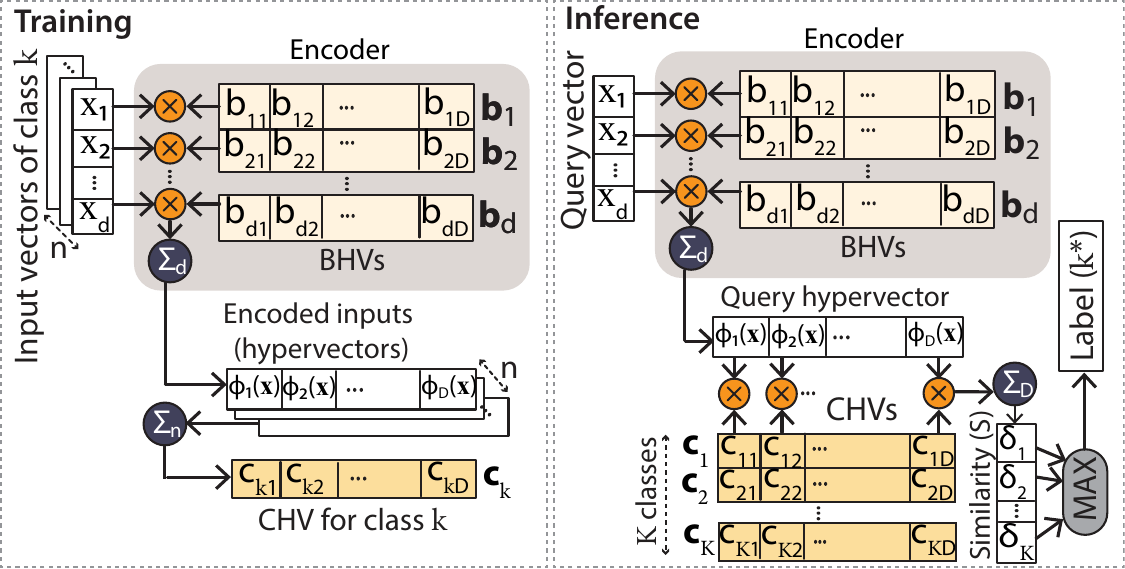}
    \caption{An illustration of the HDC training and inference processes with the \firstEnc encoding scheme.}
    \vspace{-0.15in}
    \label{fig:HDCTrainingAndInference}
\end{figure}

The training process in HDC involves mapping training data to a high-dimensional space and combining the resulting \emph{hypervectors} class-wise to generate a prototypical representative class hypervector (CHV) for each class. 
In Figure \ref{fig:HDCTrainingAndInference} (left), we show an overview of the training process with \firstEnc encoding.
Throughout this subsection, we will assume that the classification problem at hand comprises $K$ classes and we want to generate the CHV of class $k$, where $k\in[K]$.
We can generate the CHV of class $k$ in two steps: Encoding and Bundling.

\begin{figure}[t]
    \centering
     \includegraphics[width=0.43\textwidth]{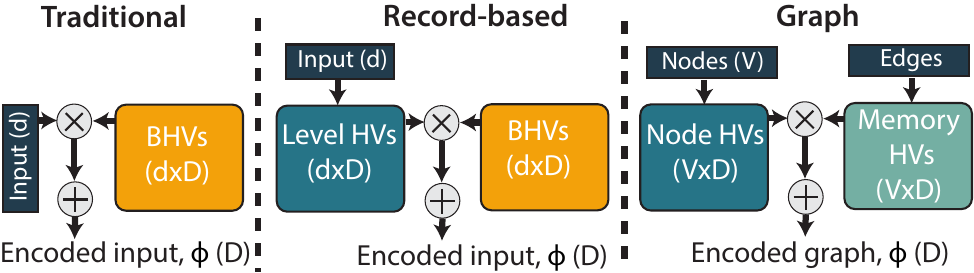}
    \caption{An overview of the three HDC encoding schemes: \firstEnc, record-based, and graph encoding.}
    \vspace{-0.25in}
    \label{fig:hdc_encodings}
\end{figure}

\noindent{\textbf{Encoding (\firstEnc)}:}
In this step, we map all the input vectors in the training dataset that belong to class $k$ from a low-dimensional space $\mathcal{X}\subset\mathbb{R}^d$ to a high
-dimensional space $\mathcal{H}\subset\mathbb{R}^D$, using a (typically \emph{randomized}) function $\phi:\mathcal{X}\to\mathcal{H}.$ 
The top half of Figure \ref{fig:HDCTrainingAndInference} (left side) illustrates this process. For example, given an input vector $\mathbf{x}\in\mathcal{X},$ and IID (Independent and Identically Distributed) sampled vectors $\{\mathbf{b}_1,\ldots,\mathbf{b}_d\}$ from $\mathcal{H}=\{\pm1\}^D$, the encoder $\phi$ is defined as
\vspace{-0.1in}
\begin{equation}
    \phi_j(\mathbf{x}) = \sum_{i=1}^{d} b_{ij} x_i,\ 1\leq j\leq D.\label{eqn:HDCencodingMVM1}
    \vspace{-0.05in}
\end{equation}

where $b_{ij}$ is the $j$\textsuperscript{th} element of $\mathbf{b}_i$. The same equation can be expressed in the matrix form as
\vspace{-0.05in}
\begin{equation}
    \phi(\mathbf{x}) = B^T\mathbf{x},   
    \label{eqn:HDCencodingMVM2} 
\end{equation}

where $B_{(d\times D)}=[\mathbf{b}_1,\ldots,\mathbf{b}_d]^T$ is the encoding matrix and $\mathbf{x}=[x_1,\ldots,x_d]^T$ is the input vector.
The vectors $\{\mathbf{b}_1,\ldots,\mathbf{b}_d\}$ are the called base hypervectors (BHVs).
In other words, we multiply each feature of the input vector with a corresponding BHV and add the results element-wise to generate the input hypervector.
Equation \eqref{eqn:HDCencodingMVM2} shows that
the encoding operation can be represented as an MVM between the encoding matrix $B_{(d\times D)}$ and the input vector $\mathbf{x}_{(d\times 1)}$.

\noindent{\textbf{Bundling:}}
After the encoding step is completed, assume that we have $n$ input vectors for class $k$, encoded to $n$ input hypervectors using Equation  \eqref{eqn:HDCencodingMVM2}.
We add the $n$ input hypervectors element-wise to generate the CHV of class $k$.
In HDC parlance, such element-wise addition operations are called \enquote{Bundling}~\cite{rosing21theoryHDC}.
After bundling, we normalize the result to fit the CHV elements within the desired bit size.

We follow this two-step process for each class in $K$ to generate all the CHVs.
This training scheme is called \enquote{single-pass} training.
Formally, if the training dataset $\mathcal{D}_{tr}$ is partitioned class-wise as $\mathcal{D}_{tr}=\bigcup_{k=1}^K\mathcal{D}_k$, then in single-pass training, the CHVs $\{\mathbf{c}_1,\ldots,\mathbf{c}_K\}\in\mathbb{R}^D$ are generated using
\begin{eqnarray}
\mathbf{c}_k := \sum_{\mathbf{x}\in\mathcal{D}_k}\mathbf{\phi(x)},~k\in[K].
\label{eqn:HDCClassHypervectorGeneration}
\end{eqnarray}
\vspace{-0.1in}

Single-pass training is the fundamental and most commonly used training method.
Specialized approaches like online or iterative training \cite{hernandez2021onlinehd} involve executing single-pass training iteratively, updating the CHVs in each iteration.  
Therefore, in our experiments, we show the results for single pass training, which provide reliable benchmarks for implementing other training methods, given that our system does not demand computationally intensive tasks like SVD for updating the CHVs.

\subsubsection{Inference}
Once training is complete, we can classify a query input by first encoding it to a $D$-dimensional space using the same BHVs used in training, then calculating similarity with each CHV generated by the training step (shown in Figure \ref{fig:HDCTrainingAndInference} (right)).
The class with the highest similarity score indicates the label for the query input.
There can be different types of similarity metrics such as the inner product, cosine similarity, and Hamming distance.
We use the widely adopted cosine similarity metric for its ability to provide better accuracy than other similarity metrics~\cite{liu2022cosime}.
Specifically, inference is performed through the following two steps:

\noindent{\textbf{Encoding}:} The query input vector $\mathbf{x}\in\mathcal{X}$ (whose class we do not know) is encoded to its corresponding query hypervector $\phi(\mathbf{x})\in\mathcal{H}$.
This encoding step is the same as the encoding done during training.

\noindent{\textbf{Similarity Measurement}:} Using a \emph{similarity metric} $\delta:\mathcal{H}\times\mathcal{H}\to\mathbb{R}$, we determine $k^*$, the class with the most similar CHV.
This $k^*$ class is determined as follows:
\vspace{-0.05in}
\begin{eqnarray}    k^*:=\argmax_{k\in[K]}~\delta\left(\phi(\mathbf{x}),\mathbf{c}_k\right)
\end{eqnarray}\vspace{-0.1in}

As mentioned earlier, we use the cosine similarity as our similarity metric.
The similarity measurement step can also be represented in a matrix form. Let $C :=[\mathbf{c}_1,\ldots,\mathbf{c}_K]^T_{(K\times D)}$ and $\mathbf{s} = [\delta_1, \delta_2, \ldots, \delta_K]^T_{K\times1}$. Then, all $K$ similarities can be simultaneously generated as
\vspace{-0.05in}
\begin{equation}
\label{eqn:mvm_sim}
\mathbf{s} = C\phi(\mathbf{x})
\end{equation}
\vspace{-0.2in}

Equations \eqref{eqn:HDCencodingMVM2} and \eqref{eqn:mvm_sim} show that encoding and similarity checking, the two primary operations in HDC, can both be represented as an MVM operation.
Note that this simplification is possible only for the chosen encoder ($\phi(\cdot)$) and the similarity metric ($\delta(\cdot)$). 


\subsubsection{Other Encoding Schemes}
We support two other encoding schemes---record-based and graph, besides the traditional encoding scheme. 
The basic flow of the procedure still remains the same for both training and inference, except for the encoding method. 
Figure~\ref{fig:hdc_encodings} succinctly illustrates the differences in the three encoding schemes.

\noindent{\textbf{Record-based Encoding}:} This encoding scheme employs two types of hypervectors, representing the feature position (\emph{position} hypervectors) and the value (\emph{value} hypervector) of the said feature \cite{ge2020hdc-lassification-review}. 
The $d$ position hypervectors $\{\mathbf{b}_1,\ldots,\mathbf{b}_d\}$ are generated IID to encode the feature position information in a feature vector. 
The feature value information is then quantized to $m$ level hypervectors $\{\mathbf{L}_1,\ldots,\mathbf{L}_m\}.$ 
Suppose the level hypervectors associated with a given input $\mathbf{x}$ are $\mathbf{L}_{1,x},\ldots,\mathbf{L}_{d,x}$ (where $\mathbf{L}_{i,x}\in \{\mathbf{L}_1,\ldots,\mathbf{L}_m\},1\leq i\leq d$), then the final encoding hypervector for input $\mathbf{x}$ is given by
\vspace{-0.05in}
\begin{equation}
    \mathcal{H}(\mathbf{x}) = \sum_{i=1}^d \mathbf{L}_{i,x}\oplus\mathbf{b}_i,
    \label{eqn:}
    \vspace{-0.05in}
\end{equation}
where $\oplus$ represents element-wise \texttt{XOR}, which is replaced with multiplication in case of multi-bits.

\noindent{\textbf{Graph Encoding}:} This encoding technique is specific to problems such as social network analysis, genomics, and knowledge graph representations, where the input data structure is a graph \cite{poduval2022graphd}. 
The challenge here is to associate with an input graph $G=(\mathcal{V},\mathcal{E})$ a hypervector that encodes the structure of the graph, i.e., the information within $\mathcal{E}$. 
Towards this end, a set of hypervectors $\{\mathbf{b}_1,\ldots,\mathbf{b}_V\}$, called node hypervectors, are generated IID, where $V=|\mathcal{V}|$. 
Next, with every node $i\in\mathcal{V},$ another hypervector $\mathbf{m}_i$, called memory hypervector, is associated that encodes the neighborhood of node $i$. 
So, for every $i\in\mathcal{V},$ $\mathbf{m}_i = \sum_{j:(i,j)\in\mathcal{E}}\mathbf{b}_j.$ 
A bundling operation then generates the encoding hypervector for this graph as follows
\vspace{-0.05in}
\begin{equation}
    \mathcal{H}(G) = \frac{1}{2}\sum_{i\in\mathcal{V}} \mathbf{b}_i\oplus\mathbf{m}_i,
    \vspace{-0.05in}
\end{equation}
This basic technique can be extended in a fairly straightforward manner to incorporate edge weights and directed graphs. 
For details, we refer the reader to the GrapHD work\cite{poduval2022graphd}.
Note that the graph encoding scheme follows the same process as record-based encoding after we have generated the node hypervectors and the memory hypervectors.


\begin{figure}[tb]
    \centering
    \includegraphics[width=0.4\textwidth]{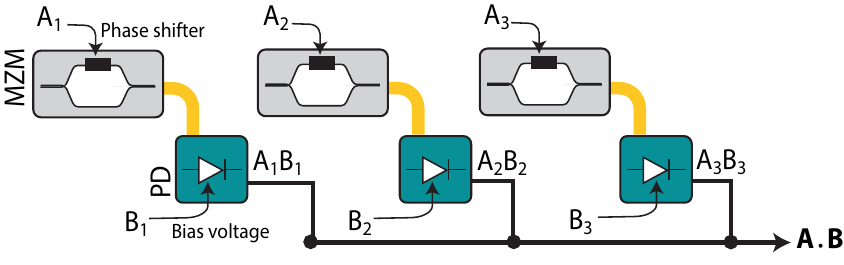}
    \vspace{-0.1in}
    \caption{Dot product of vectors $\mathbf{A}$ and $\mathbf{B}$ using MZMs and PDs.}
    \vspace{-0.25in}
    \label{fig:mzm_pd_dot}
\end{figure}

\subsection{Photonic Computing Overview}\label{subsec:background_photonics}

Two prominent types of electro-photonic accelerators for DNN inference: MZI array-based~\cite{cansu2021electro}\cite{shen2017deep} and MRR weight bank-based~\cite{deap-cnn}\cite{sunny2021crosslight} accelerators, are capable of running HDC training and inference \emph{with traditional encoding}, where one of the operands can be kept stationary in the photonic array (MZI array or MRR weight bank).
However, MZI arrays are unsuitable for iterative training due to the need for SVD on operands before being programmed in the MZI arrays.
For the other encoding schemes, both MZI and MRR-based accelerators demand high energy consumption.
In record-based and graph encoding schemes, element-wise multiplications are performed between level hypervectors and BHVs, which requires $O(N^2)$ data modulation per cycle in an $N\times N$ optical unit.

To address these challenges, in our work, we utilize a MAC unit based on an MZM and PD pair to perform vector dot products.
An MZM can modulate the amplitude of the optical input signal through attenuation.
The amount of attenuation can be controlled by changing the applied voltage to the MZM. 
MZMs can be modulated with tens of GHZ rates through the free carrier dispersion effect-based actuation mechanisms~\cite{ps_opt_electrical}
%
A PD is a light-sensitive diode that generates current (called photocurrent) when exposed to optical signals of sufficient power.
The photocurrent is proportional to the power of the incident optical signal.
By applying a programmable bias voltage across a PD, one can control the generated maximum photocurrent \cite{PD_sheng2010ingaas}\cite{li2022spacx}.
In other words, at the output of a PD, we can obtain a photocurrent that is proportional to the product of the optical signal power and the applied voltage, which effectively performs a multiplication operation. 

Figure \ref{fig:mzm_pd_dot} shows how we can use MZMs and PDs to perform a dot product between two vectors $\mathbf{A} = [A_1, A_2, A_3]$ and $\mathbf{B} = [B_1, B_2, B_3]$.
First, we modulate the light waves through three MZMs by controlling the voltages applied to the phase shifters corresponding to $A_1$, $A_2$, and $A_3$.
We pass the modulated light waves through PDs, of which we control the bias voltages corresponding to $B_1$, $B_2$, and $B_3$.
Therefore, we obtain currents representing $A_1 \cdot B_1$, $A_2 \cdot B_2$, and $A_3 \cdot B_3$ at the output of the PDs.
We direct the currents into a single wire to accumulate the multiplication results, and thus, perform a dot product.
This basic \emph{device-level idea} of utilizing MZMs and PDs for performing Multiply-Accumulate (MAC) operations was first proposed by Harris \cite{harris2021photonics}. 
In this paper, we build the first-ever all-in-one comprehensive HDC accelerator using this MZM-PD-based device-level idea as the fundamental building block.

\section{Electro-Photonic Accelerator for HDC}\label{sec:accel_microarch}
\begin{figure*}[t]
    \centering
    \includegraphics[width=0.75\textwidth]{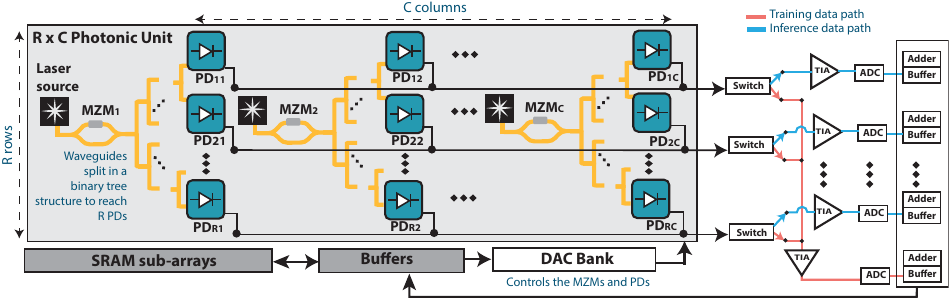}
    \caption{Overview of the \acceleratorname accelerator microarchitecture consisting of an \rxc photonic unit, data converters, SRAM-based memory units, and other circuits. The photonic unit uses one MZM per column to modulate the BHVs or CHVs and a PD array of $R$ rows and $C$ columns to perform multiplication with corresponding input values.}
    \vspace{-0.15in}
    \label{fig:arch_main}
\end{figure*}
This section delves into the microarchitectural details of our proposed accelerator, \acceleratorname.
We first introduce the high-level design of the \acceleratorname microarchitecture and discuss the individual components in detail. 
We then explain the dataflow for performing HDC training and inference in \acceleratorname.

\subsection{\acceleratorname Microarchitecture}
\label{subsec:arch_components}
Figure \ref{fig:arch_main} presents an overview of the \acceleratorname accelerator comprising an \rxc photonic unit, data converters, switches, and adders.
The photonic unit contains PDs and MZMs to perform MAC operations.
The SRAM-based memory units store the BHVs, CHVs, inputs, intermediate results, and final results.
We provide a detailed description of each component below.

\subsubsection{Lasers}
The optical signals are generated using off-chip lasers and guided by a single-mode fiber.
There are $C$ lasers in an \rxc photonic unit.
The light from each laser is coupled to a waveguide via couplers and fed to an MZM.
\subsubsection{MZMs}
An \rxc unit contains $C$ MZMs, one for each column.
These MZMs are responsible for modulating the optical signals to represent the values in BHVs or CHVs, depending on the ongoing operation (training or inference with \firstEnc or record-based encoding).
The modulated optical signal from each MZM is propagated to all rows in the corresponding column via $\lceil\text{log}_2R\rceil$ cascaded splitters.
The columns are optically isolated from each other.

\subsubsection{PD Array}
The \rxc photonic unit in Figure \ref{fig:arch_main} contains an \rxc array of PDs, i.e., a total of $R \cdot C$ PDs.
The bias voltages of the PDs are adjusted based on the operands that are required for the MAC operation with the BHVs or CHVs.
These operands can be input feature values, level hypervectors, or encoded data depending on the accelerator operation. 
As the sizes of these matrices are typically larger than the \rxc photonic unit, we partition the matrices into tiles of size \rxc and program them to the PD array one tile at a time.
After the PDs are programmed, each PD receives the modulated optical signal from the MZM of the corresponding column and generates a photocurrent proportional to the product of the optical power it receives and the applied bias voltage, effectively performing a multiplication operation between the values programmed into the MZM and the PD.
The output currents (multiplication results) of the PDs in a row are accumulated by a wire, which is connected to a switch that routes the current to the appropriate destination, based on the accelerator operation.

\subsubsection{Switches}
In \acceleratorname, we use switches (one switch per row) to either forward the current to the associated Trans-Impedance Amplifier (TIA) or to the wire that accumulates the currents coming from all the rows.
This accumulation helps us efficiently perform bundling in training, which will described in detail in Section~\ref{subsec:arch_functionality}.

\subsubsection{Data Converters}
As described in Section \ref{subsec:background_photonics}, the MZMs and the PDs are programmed or controlled by analog voltages.
Therefore, the operands, which are stored in the SRAM arrays as digital values, need to be converted into the analog domain using Digital-to-Analog Converters (DACs).
An \rxc unit requires $C$ DACs for the MZMs and \rxc DACs for the PD array.

At the end of each row, the received currents carry the MAC results which are converted into voltages through TIAs and then converted back to the digital domain through Analog-to-Digital Converters (ADCs).
In an \rxc unit, we have $R+1$ TIAs and $R+1$ ADCs, one for each row and one dedicated to the bundling output.

\subsubsection{Adder-Buffer Units}
After each ADC, we include CMOS-based digital adders and buffers to accumulate the partial results when necessary.
To match the throughput of the photonic unit, we use $f$ adders in each row for a photonic unit operating at $f$ GHz,  where each adder can perform addition within 1 ns with $1/f$ ns offset to one another.
Typical value for $f$ is 5-10 GHz.
The buffers included with the adders hold the operands and the adders' outputs.

\subsubsection{Memory Units}
We use SRAM arrays to store the BHVs, CHVs, and input vectors.
We use a double-buffering approach for storing the input vectors where we perform operations on one batch of input while transferring the next one to avoid data movement delays.  
In the case of record-based or graph encoding, we store the level hypervectors in an SRAM look-up table.
The total required size of the SRAM-based on-chip memory is $<1.4$ MB for our datasets.

\subsection{\acceleratorname Functionality and Dataflow}
\label{subsec:arch_functionality}
\begin{figure*}[t]
        \centering
        \begin{subfigure}[b]{0.41\textwidth}
            \centering
            \includegraphics[width=\textwidth]{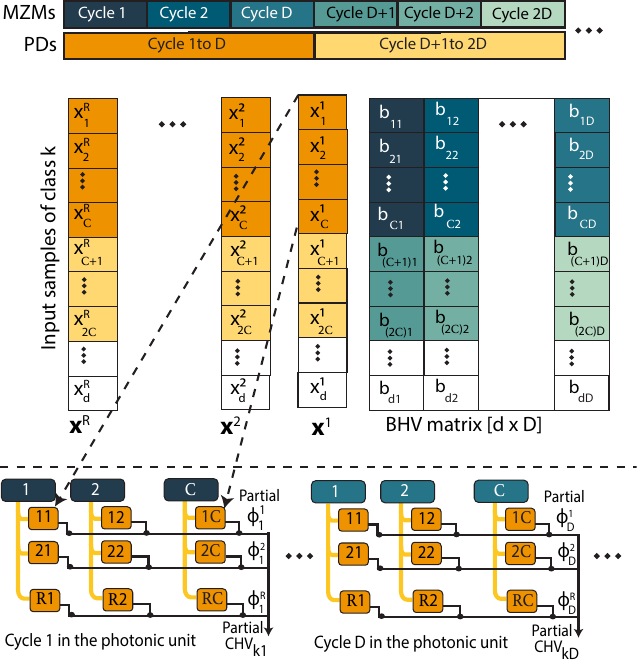}
            \vspace{-0.2in}
            \caption{Training with \firstEnc encoding}   
            \label{fig:dataflow-train-basic}
        \end{subfigure}
        \hspace{0.3in}
        \begin{subfigure}[b]{0.41\textwidth}  
            \centering 
            \includegraphics[width=\textwidth]{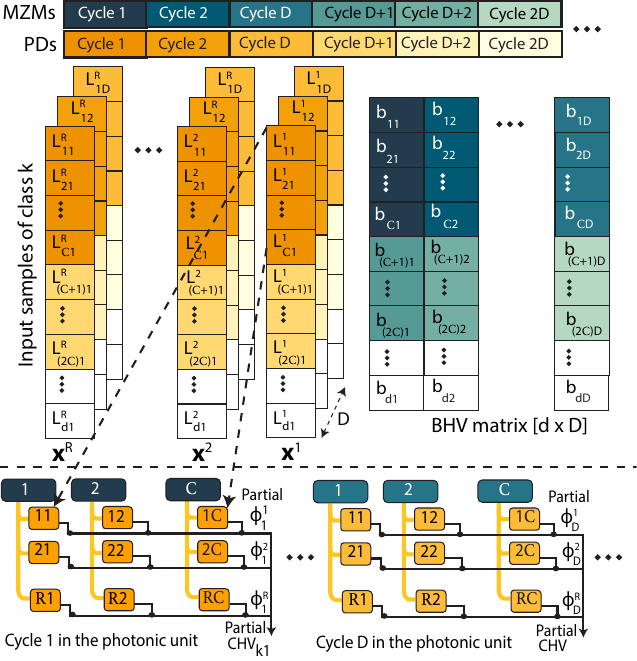}
            \vspace{-0.2in}
            \caption{Training with record-based/graph encoding}     
            \label{fig:dataflow-train-record}
        \end{subfigure}
        \vspace{-0.05in}
        \caption{Dataflow of HDC training with \acceleratorname for \firstEnc and record-based/graph encoding methods (figure shows the processing of the first 2C features). The PD array processes the input values tile by tile, and for each tile, the MZMs modulate the corresponding BHV portions. Partial encoding results are generated in every row and bundling is performed by summing the row currents. ($\phi_j ^i$ indicates $\phi_j(\mathbf{x}^i)$).}
        \vspace{-0.2in}
        \label{fig:dataflow-train}
\end{figure*}

We developed dataflows for running HDC training and inference with each encoding scheme on \acceleratorname.
After exploring various dataflow options (e.g., weight stationary, output stationary, etc.), we designed our dataflow primarily based on an input stationary approach, making necessary adjustments for performing different HDC operations efficiently.
Our dataflow provides efficient bundling, supports a diverse set of HDC operations, and optimizes the number of required data converters.
Below we describe the dataflows for each HDC operation in detail. 
Please note that the dataflows for graph encoding are not provided as we use the same dataflows for record-based encoding and graph encoding for both HDC training and inference. 
The required sets of operations for the two encoding schemes are the same, assuming node and memory hypervectors are pre-computed in the graph-based scheme (See Section~\ref{sec:background}).

\subsubsection{Training with Traditional Encoding} 
Figure \ref{fig:dataflow-train-basic} provides a detailed illustration of the dataflow for training with traditional encoding.
Here, we first program the \rxc PD array using $C$ feature values from $R$ samples ($\mathbf{x}^1 \cdots \mathbf{x}^R$) at a time.
Next, we employ the MZMs to modulate the corresponding $C$ elements of each hyperdimension (column) of the BHVs in every cycle.
Therefore, each row's output is the partially encoded data pertaining to the $C$ feature values of each sample.
For example, in cycle 1, we load the first \rxc input tile into the PD array, and modulate the corresponding $C$ elements of the BHV matrix, $b_{11}$ to $b_{c1}$, through the MZMs.
This generates the partial encoding of the first hyperdimension, involving the first $C$ feature values for the $R$ samples (partial $\phi^1_{1}$ to $\phi^R_{1}$).
In the next cycle, we modulate the second column of the BHV matrix, $b_{12}$ to $b_{c2}$, with the same input tile existing in the PD array, and so on.

As the encoding step is followed by a bundling operation, the outputs from different samples belonging to the same class are accumulated.
Therefore, during the encoding step in training, we control the switches to direct the photocurrents onto the \emph{training data path} (orange path in Figure \ref{fig:arch_main}) for bundling via current accumulation.
Assuming that the $R$ samples belong to class k, the summed current represents the partial CHV of class $k$.
The summed current is converted to the digital domain with an ADC and stored in the buffer.
This enables us to group $R$ samples from the same class $k$ and perform encoding and bundling together for an input tile with only one TIA and ADC.

These MVM operations continue until all the BHVs are multiplied with the same input tile and bundled. 
This takes $D$ cycles for a single tile.
The same operations are repeated for all the tiles and the partial CHVs are accumulated digitally to obtain the final CHVs and written back to SRAM. 
For input samples with $d$ features, it takes $\left\lceil d/C \right\rceil \times D$ cycles to complete the encoding and bundling of $R$ samples.
By employing multiple photonic units, the input tiles can be distributed among the photonic units and processed in parallel.

\begin{figure}[t]
    \centering
    \includegraphics[width=0.45\textwidth]{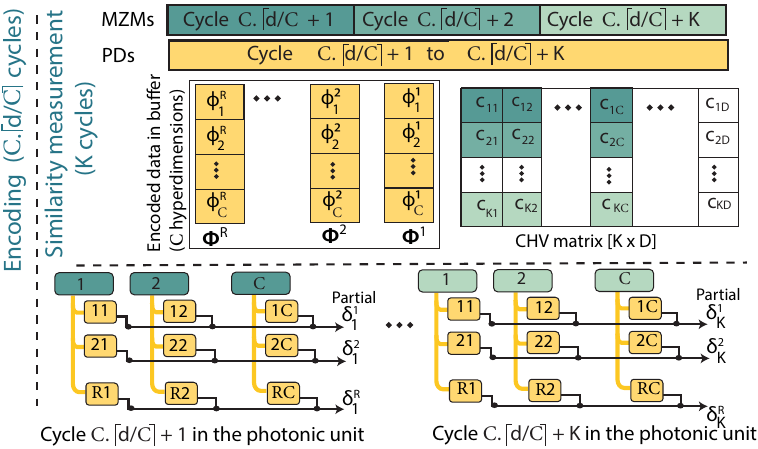}
    \vspace{-0.05in}
     \caption{Inference dataflow in \acceleratorname with \firstEnc encoding.}    
    \vspace{-0.2in}
    \label{fig:dataflow_inf}
\end{figure}

\subsubsection{Training with Record-based Encoding}
For record-based encoding, we follow a similar procedure as \firstEnc encoding.
However, unlike the \firstEnc, record-based encoding maps each $d$ feature of an input vector to a $D$-dimensional level hypervector.
Therefore, one input feature value is not shared with all the elements of the BHVs as in the \firstEnc encoding.
We demonstrate this process in Figure \ref{fig:dataflow-train-record}.
In the figure, we load the first $C$ elements of the first column of the BHVs into the MZMs and the corresponding $C$ elements from the level hypervectors of the $R$ samples into the PDs.
Different from \firstEnc encoding, we cannot keep this input tile in the PDs stationary for the next $D-1$ BHV columns as each hyperdimension corresponds to a different level hypervector.
This requires updating the values both in the MZMs and the PDs for each column of BHVs every cycle.

Updating the values in MZMs and PDs every cycle is possible, however, it requires a higher number of SRAM reads and D-A conversions compared to the \firstEnc. 
The rest of the training procedure is similar to the \firstEnc encoding during training.
The amount of computation is the same as the \firstEnc encoding and it takes the same $\left\lceil d/C \right\rceil \times D$ cycles to complete the encoding and bundling of $R$ samples.


\subsubsection{Inference with Traditional Encoding}
Figure \ref{fig:dataflow_inf} shows the inference dataflow with \acceleratorname for \firstEnc encoding.
As described in Section \ref{subsec:background_hdc}, for inference, we need to first encode the input sample, then calculate the similarity scores between the encoded sample and all the CHVs to get the class label.
The mapping strategy of encoding during inference is exactly the same as training except that we do not need to bundle the encoded samples during inference.
Therefore, we control the switch to direct the output of each row to the ADC corresponding to that row, therefore, keeping the results of each row separate (blue path in Figure \ref{fig:arch_main}).

We keep the input tile stationary in the PD array for $C$ cycles and process $C$ columns of the BHVs for the corresponding tile, similar to training.
Repeating this for all $d$ rows of the $R$ samples takes $\left\lceil d/C \right\rceil \times C$ cycles.
We store the results in an \rxc buffer at the output and accumulate the partial results as they are generated.

We use the next $K$ cycles to perform similarity measurements with the $K$ CHVs (calculated during training) for the corresponding $C$ hyperdimensions we encoded.
We load the encoded \rxc tile from the buffer to the PD array and modulate the MZMs to encode CHVs for each class one at a time.
Thus, we generate an output of shape $R\times K$, containing the partial similarity scores for all classes for the $R$ samples.
We store this result in the buffer and repeat this process for all $D$ hyperdimensions ($C$ hyperdimensions at a time).
After a total of $\left\lceil D/C \right\rceil \times ((\left\lceil d/C \right\rceil \times C) + K)$ cycles, we obtain the final similarity scores for the $R$ samples.
By calculating the maximum class scores for the R samples, we get $R$ labels and complete a batch inference.

\subsubsection{Inference with Record-based Encoding}
Inference with record-based encoding follows the same procedure as the inference with \firstEnc encoding where the encoding scheme is replaced with the record-based one which is described in Section~\ref{subsec:background_hdc} without the bundling operation.
As the rest of the procedure is the same as the inference with \firstEnc encoding, the number of cycles is again, $\left\lceil D/C \right\rceil \times ((\left\lceil d/C \right\rceil \times C) + K)$, for a batch inference of $R$ samples.


\subsection{Performance vs. DAC area tradeoff}
\label{subsec:dac_opt}
As mentioned in Section \ref{subsec:arch_components}, an \rxc photonic unit requires \rxc DACs for the PD array and $C$ DACs for the MZMs.
This large number of DACs significantly increases the area of \acceleratorname.
To reduce the area, we can share a DAC across multiple PDs in a time division multiplexed manner while loading a new tile in the PD array.
This causes a delay, $t_{DAC}$, during tile update, which does not affect the overall system performance if the tiles do not need to be updated frequently (more details about this below).
The connection between one DAC and multiple PDs can be achieved with a standard analog demultiplexer circuit~\cite{analog-demux}.
MZI-array-based accelerator like ADEPT also employs the DAC sharing method to program multiple MZI cells during the 10 ns weight settling period~\cite{cansu2021electro}.
In our case, we intentionally allow a delay to reduce the number of DACs.

For instance, in a $128\times128$ photonic unit operating at 5 GHz, if each PD is controlled by a dedicated DAC with a 5 GHz sampling rate, it would require 16,384 DACs.
Each PD would then be programmed within a single clock cycle (0.2 ns) using a dedicated DAC.
However, sharing a DAC across multiple PDs, let's say 6 PDs, reduces the required DAC count to $\lceil 16,384/6\rceil = 2,731$.
Although this introduces an additional latency overhead of $t_{DAC}$ = 1 ns between consecutive tiles (we have a 0.2 ns cycle time), this performance overhead is negligible compared to the reduction in area due to the reduction in the number of DACs.

This $t_{DAC}$ delay has a minimal impact on performance for training and inference with traditional encoding, where we do not update the tiles frequently.
However, in record-based encoding, we have to update the tile every cycle, therefore, the delay added between each operation results in a lower performance.
So we do not implement DAC-sharing for record-based encoding.
We investigate the trade-off between area improvement and performance loss for the DAC-sharing approach in detail in Section \ref{sec:results} for \firstEnc encoding.

\subsection{Integration technology for \acceleratorname}\label{subsec:integration} 
In PhotoHDC, similar to ADEPT~\cite{cansu2021electro}, we use a 3D integration approach where the electronic and photonic chiplets are integrated vertically. 
This enables us to use different technology nodes for the two chiplets.
The photonic chiplet consists of the MZM array while the electronic chiplet includes PDs, DACs, ADCs, TIAs, and SRAM arrays. 
Alternative integration options include monolithic integration~\cite{batten2009building}~\cite{giewont2019300-monolithic-gf} where all the components are fabricated on a single chip and 2.5D integration~\cite{sunny2023-2.5d-machine}~\cite{li2022spacx} where chiplets of individual components are integrated next to each other on an interposer.
The broader conclusions in our paper are applicable even if we design PhotoHDC using monolithic or 2.5D integration.

\section{Evaluation Methodology}\label{sec:eval_method}
In this section, we provide a detailed description of the methodology that we used for calculating the Power, Performance, and Area (PPA) of \acceleratorname.
We also describe how we mapped HDC training and inference in three SOTA electro-photonic DNN accelerators, and CiM-based designs of HDC accelerators for comparison.

\subsection{PPA Estimation of \acceleratorname}\label{subsec:ppa_model}
\subsubsection{Power/Energy Estimation}\label{subsubsec:power_estimation}
In \acceleratorname, power is consumed by SRAM arrays, adders, data converters, Electrical-to-Optical (EO) and Optical-to-Electrical (OE) converters, and the laser.
We generate the SRAM arrays with an SRAM compiler for the GF22FDX technology node~\cite{22fdx} to estimate the energy consumption per access. 
We count the total number of SRAM reads and writes, and calculate the SRAM energy. 
For addition operation in \acceleratornametrain, we designed a 32-bit adder in RTL and synthesized it using Genus \cite{genus} with the GF22FDX technology node.

The laser power is estimated by calculating the signal-to-noise ratio (SNR) required to guarantee the desired bit precision at the PD output accounting for all the losses on the optical path.
To ensure $b$-bit precision, an $\text{SNR}\geq2^b$ should be achieved at the detection.
The optical signal power at a PD needs to be $(\kappa \;\text{SNR}_{\text{shot}})^2  \cdot (q \; \Delta f/4)$, where $\text{SNR}_{\text{shot}}$ is the SNR with respect to the shot noise, $\kappa$ accounts for all other noise contributions---thermal noise, $q$ is the elementary charge, and $\Delta f$ is the bandwidth of the detector \cite{cansu2021electro}.
We consider a noise factor of 3~\cite{cansu2021electro}, a fiber-to-chip coupling loss of 2 dB~\cite{cansu2021electro}, a loss at the MZMs of 1.2 dB~\cite{MZM_Akiyama:12}, splitting loss of 0.2 dB~\cite{batten2009building} for each split (total $\log_2R$ splits), waveguide loss of 1.5 dB/cm for straight and 3.8 dB/cm for each bend ~\cite{waveguideloss-chrostowski2016design} (with a bend radius of 0.5 $\mu m$~\cite{sakai2001sharplybent}), PD responsivity 1.1 A/W~\cite{PD_sheng2010ingaas}, and a laser efficiency of 20\%~\cite{soton356442}.
%
%
%
The modulation with the MZMs consumes $\sim$20 fJ/bit~\cite{sun2015single}, and the tuning circuits of the MZMs consume 11.3mW~\cite{MZM_Akiyama:12}.
TIAs consume 75 fJ/bit~\cite{TIA_rakowski2018hybrid}.

Using the energy numbers of a 14-bit DAC\cite{dac_huang202110} and a 10-bit ADC\cite{adc_guo201929mw} of the 28 nm technology node, we apply bit-precision scaling to estimate the energy consumption of the lower precision output ADCs, input DACs, and weight DACs.
Both DAC and ADC energy consumption scales roughly with $2^\text{ENOB}$ where ENOB is the effective number of bits of the converter. 
We use this approach as there are no publicly available DAC and ADC implementations or prototypes in GF22FDX with our desired bit precision and throughput.

\subsubsection{Latency Estimation}\label{subsubsec:latency_estimation}
In Section \ref{subsec:arch_functionality}, we describe each dataflow in detail and the corresponding cycle counts for R number of samples.
Using those cycle counts and the accelerator frequency (maximum 5 GHz), we calculate the total training latency considering all training samples and the inference latency for the total number of inferences.
Both MZMs~\cite{MZM_Akiyama:12} and PDs~\cite{PD_sheng2010ingaas} have modulation bandwidth $\geq5$ GHz, therefore, they do not cause any latency overhead during operations.

\subsubsection{Area Estimation}
For SRAM and the adders, we estimate the area for the GF22FDX technology node~\cite{22fdx}.
The area of the abovementioned ADCs and DACs are scaled down to 22 nm technology node assuming a fixed voltage scaling.
Each MZM takes up $300\times50\ \mu m^2$~\cite{MZM_Akiyama:12} and the area of a PD is $40\times40\ \mu m^2$~\cite{PD_sheng2010ingaas}.

\subsection{Behavioral Simulation and System Robustness}\label{subsec:behav} 
To verify the functionality of the photonic product unit we simulated it with the Virtuoso Spectre platform using VerilogA. 
Our simulation results verify that the dot product unit can successfully operate at 5 GHz with up to 10-bit precision.

In addition, the short optical path in our design, unlike the prior works with multiple cascaded devices~\cite{cansu2021electro}\cite{sunny2021crosslight}, minimizes the error accumulation and enables us to easily calibrate away any phase offset that might be caused by process variations, fabrication errors, and thermal fluctuations. 
This phase offset can be eliminated by measuring the phase at the coherent detection without any voltage applied on the phase shifters and adjusting the applied voltage range accordingly~\cite{harris2021photonics}. 
This calibration can be repeated over time to get rid of the unwanted phase bias due to environmental changes.

\subsection{HDC with SOTA Electro-Photonic DNN Accelerators}
\label{subsec:sota_mapping}
Below we describe how we map HDC training and inference operations to SOTA electro-photonic DNN accelerators.
This methodology will be used in Section~\ref{sec:results} for providing a quantitative comparison against \acceleratorname. 
Traditional encoding and similarity measurement are both represented as MVM operations and therefore, can be mapped to a DNN accelerator similar to a fully-connected (FC) layer in a DNN accelerator.
We only consider the \firstEnc encoding for SOTA electro-photonic DNN accelerators as other encoding schemes require modulating  O($N^2$) devices every cycle, which is highly inefficient in these accelerators tailored for DNNs. 

\subsubsection{ADEPT~\cite{cansu2021electro}}
For training with \firstEnc encoding on ADEPT, we perform encoding and bundling in a two-stage pipelined manner.
We utilize the MZI arrays to encode the input data through an MVM of the BHVs and the input vector, with the BHVs as weights within the MZI array.
Then, we bundle the encoded inputs using adders of the vector processing units in ADEPT.
For inference, we first encode the input samples similar to training, then perform MVMs between the CHVs (weights) and the encoded input to calculate similarity scores.
We use a 128$\times$128 MZI array (two arrays for inference) as suggested in the ADEPT paper for both encoding and similarity measurement.


\subsubsection{ALBIREO~\cite{shiflett2021albireo}}
In ALBIREO, for performing HDC operations, we follow the same dataflow described in the paper for executing FC layers.
We use the \emph{``conservative''} version of Albireo with 60 W power consumption.
Here too, we assume that there are two similar cores for inference, and the encoding and similarity measurement steps are pipelined.

\subsubsection{DEAP-CNN~\cite{deap-cnn}}

We perform \firstEnc encoding in DEAP-CNN by modulating the BHVs in the kernel bank and the inputs in the input bank to perform dot product between them.
For inference, we calculate the similarity scores by loading the CHVs in the kernel bank and the encoded input in the input bank in a similar manner.
We follow the same pipelining assumption as the previous two accelerators.
We use the 1024 MRR design with two units that consume 112 W power as mentioned in the paper. 


\vspace{-0.05in}
\subsection{HDC with CiM-based Acclerator}\label{subsec:cim_neurosim}
We also compare the performance of \acceleratorname with CiM-based accelerator for traditional encoding.
We modify DNN-NeuroSim \cite{peng2019dnn+,peng2020dnn+} for estimating hardware performance for ReRAM crossbar arrays when running HDC training and inference.
We represent the HDC encoding and similarity measurement as FC layers as in DNNs and modify the network configuration file in the simulator accordingly.
Specifically, we use DNN-NeuroSIM v1.3 to simulate encoding and cosine similarity measurement, and DNN-NeuroSIM v2.1 to simulate the CHV update on the CiM platform.
Note that DNN-NeuroSIM v2.1 generates the results for the complete training of a DNN layer including the forward pass, error calculation, gradient calculation, and weight update.
We only consider the weight update latency and energy to calculate the total cost of the CHV update.

In DNN-NeuroSIM, we use a 22nm technology node similar to \acceleratorname for a fair comparison.
For the ReRAM crossbar array, we use a 128$\times$128 subarray size and Successive Approximation Register (SAR)-ADCs.
The chosen 128$\times$128 size has been shown to achieve relatively balanced trade-offs among accuracy, energy efficiency, throughput, area, and memory utilization \cite{peng2019dnn+}.
For CiM too, we follow the same pipelining strategy as the SOTA electro-photonic accelerators in training and inference.

\subsection{Datasets, Hyperdimension, and Bit-Precision}\label{subsec:datasets}

\renewcommand{\tabcolsep}{2.2pt}
\renewcommand{\arraystretch}{0.8}
\begin{table}[tb]
\caption{Datasets used for evaluation.}
\vspace{-0.08in}
\label{tab:datasets}
\begin{tabular}{lllll}
\hline
\textbf{Dataset} & \textbf{d}             & \textbf{K} & \textbf{Train size} & \textbf{Description}              \\ \hline
ISOLET~\cite{isolet-Dua:2019}          & 617                    & 26         & 6238                & Alphabet recognition from voice   \\
UCIHAR~\cite{ucihar_anguita2013public}           & 561                    & 12         & 6231                & Human activity recognition        \\
FACE~\cite{face-angelova2005pruning}             & 608                    & 2          & 522441              & Gender detection from images \\
PAMAP~\cite{pamap-reiss2012introducing}            &  75                    & 5          &  611142         & Activity Recognition (IMU)        \\
PECAN~\cite{pecan}            &  312                  &  3          &  22290                 & Urban Electricity Prediction      \\ \hline
                 & \textbf{Avg. \#} & \textbf{K} & \textbf{Train size} & \textbf{Description}              \\
                 & \textbf{vertices} & & &              \\ \hline
DD~\cite{dd-protein}               & 285                    & 2          & 1178                & Classify proteins                 \\
ENZYMES~\cite{enzymes}          & 33                     & 6          & 600                 & Classify enzymes                  \\
PROTEINS~\cite{dd-protein}         & 40                     & 2          & 1113                & Classify proteins                 \\ \hline
\end{tabular}
\vspace{-0.2in}
\end{table}

Table~\ref{tab:datasets} shows the list of the datasets we used for evaluation.
The columns $d$ and $K$ refer to the number of input features and the number of classes, respectively.
The train size column represents the number of samples used for training, which is chosen to be $70\%$ of the original dataset size.
For \firstEnc and record-based encoding, we use the first five datasets.
The bottom three are graph datasets and we use them for graph encoding.
We choose these datasets due to their popularity in HDC research~\cite{hernandez2021onlinehd}\cite{kazemi2021mimhd}\cite{nunes2022graphhd}\cite{kim-date-2023efficient}\cite{imani21scalableHDC}, and their diversity.
We use 4-bit precision with $D$ = 4096 in all our experiments.
Although many popular HDC implementations use binary models with higher $D$ due to the hardware friendliness of the binary operations, binary models cannot achieve high accuracy in complex applications. 
Therefore, to demonstrate the efficient multi-bit operation capability of photonics, we use the above-mentioned bit-precision and $D$~\cite{hernandez2021onlinehd}\cite{ge2020hdc-lassification-review}\cite{rahimi2016robust}\cite{kim-date-2023efficient}\cite{kazemi2021mimhd}.

\section{Evaluation Results}\label{sec:results}




\subsection{Hardware Parameter Selection in \acceleratorname}
We begin our analysis first by exploring how the choice of different parameters in \acceleratorname affects power, performance, and area.
We then conduct an exhaustive search to identify the optimal hardware parameters in \acceleratorname for HDC training and inference with all encoding schemes. 

\begin{figure}[tb]
    \centering
    \includegraphics[width=0.44\textwidth]{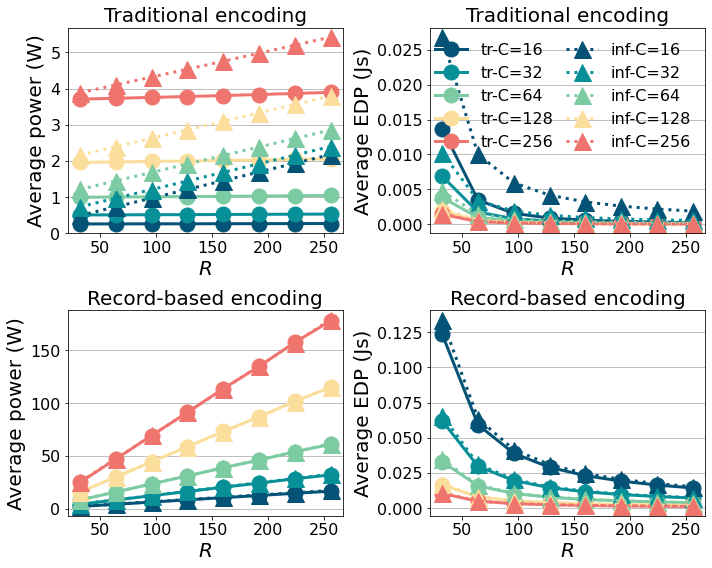}
    \vspace{-0.1in}
    \caption{Power (left) and EDP (right) trends as we increase the size of the photonic unit for training (tr) and inference (inf).}
    \vspace{-0.15in}
    \label{fig:scalability}
\end{figure}

\begin{figure}[t]
    \centering
    \begin{subfigure}[t]{0.23\textwidth}
             \centering
             \includegraphics[width=\textwidth]{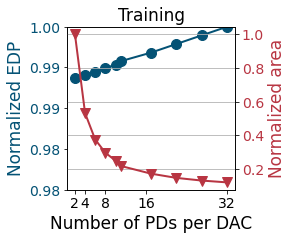}
             \label{fig:delay_train}
         \end{subfigure}
         \begin{subfigure}[t]{0.23\textwidth}
             \centering
             \includegraphics[width=\textwidth]{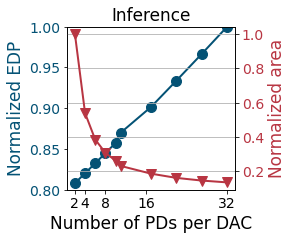}
             \label{fig:delay_inf}
         \end{subfigure}
    \vspace{-0.22in}
    \caption{Average EDP and the corresponding area for different \#PDs/DAC on a 128$\times$128 core running at 5 GHz in the \firstEnc encoding scheme. Enforcing DAC sharing leads to a slight rise in EDP but a significant reduction in area, particularly during training.}
    \vspace{-0.2in}
    \label{fig:dac_delay}
\end{figure}

\subsubsection{Impact of $R$ and $C$ on Power and EDP (Scalability)}
In Figure \ref{fig:scalability}, we show the impact of changing $R$ and $C$ on average power consumption (left) and EDP (right) in \acceleratorname across the evaluated datasets.
Here, the accelerator is running at 5 GHz with one photonic unit as an example.

On the left-side plots ((a) and (c)), the key trend to notice is that the power consumption increases linearly with the size of the photonic unit.
This is because each optical channel involves only one MZM and one PD on the optical path, regardless of the size of the photonic unit. 
This linear relationship results in improved scalability compared to typical photonic accelerators where optical devices are cascaded and the power increases exponentially with the array size. 
The power consumption for inference increases faster than training due to the difference in ADC usage: in inference, we employ one ADC per row, thus resulting in a linear power increase with $R$, whereas for training, only one ADC is used for the whole photonic unit (See the inference and training paths in Figure~\ref{fig:arch_main}).

Second, we observe that it is better to scale up along the rows (design a taller core) than the columns (design a wider core).
This is because increasing the number of columns increases the number of MZMs in the design and results in extra power consumption for MZM tuning---which can dominate the overall power consumption.
Therefore, increasing the number of rows enables broadcasting the output of a single MZM to more PDs and is a more efficient scaling-up option than increasing the number of columns.

Finally, as we increase the size of the optical units, latency decreases at a faster rate than power increases. 
This leads to lower energy consumption and EDP for larger photonic unit sizes as long as the resources are fully utilized given a dataset (Figure~\ref{fig:scalability}(b) and (d)).

When deciding the photonic unit size, one must also consider the total wire delay from one end of the unit to the other and ensure that the data travels fast enough to be compatible with the clock rate of the system.
The maximum length of the electrical wires at the output of the PDs is $\approx$1.2 cm in a 128$\times$128 photonic unit, which is the maximum size in our design space exploration (Section \ref{subsec:best_edap}).
This maximum wire length in a row causes a delay of $\approx$0.04 ns~\cite{copper-permeability}, which is well within 5 GHz operating frequency (0.2 ns cycle time) and does not cause a latency overhead.




\subsubsection{Impact of DAC sharing on Area and EDP}\label{subsec:tdac_analysis}

Next, we investigate the impact of DAC sharing on EDP and area in \acceleratorname for \firstEnc encoding.
Figure \ref{fig:dac_delay} shows the average EDP and area across the datasets for a 128$\times$128 photonic unit running at 5 GHz.
Both area and EDP are normalized with respect to the maximum data point.
The DACs we use~\cite{dac_huang202110} have a sampling rate of 10 GS/s, so in one clock cycle (0.2 ns), one DAC can perform two conversions to program two PDs without any latency overhead.

On the left plot in Figure \ref{fig:dac_delay}, we observe that even by sharing a DAC with a small number of PDs, as few as 8, we can achieve a  $\approx$70\% reduction in the total area with only $<$1\% increase in EDP.
This EDP increase is minimal because the latency overhead is only introduced when the tile is updated.
For example, the extra delay $t_{DAC}$ caused by sharing one DAC with four PDs is only one cycle and is introduced once in $D=4096$ cycles during training, which is insignificant. 
As a result, this approach enables us to halve the number of DACs we use in the system and improve the area efficiency with a negligible increase in latency. 
Similarly, for inference, this DAC sharing approach reduces the area by $\approx$70\% with only $\approx$5\% increase in EDP when one DAC is shared with 8 PDs.

\subsubsection{Energy-Delay-Area Product (EDAP)-optimized design}\label{subsec:best_edap}
We explore the design space for each encoding scheme through an exhaustive search and identify the architectural parameters that minimize the average EDAP across all the datasets that we have considered, under a 20 W power and 500 mm$^2$ area budget.
We list the parameters and their latency and power consumption in Table \ref{tab:best_params}.
We use these parameters throughout the rest of the experiments.

\renewcommand{\tabcolsep}{2.3pt}
\renewcommand{\arraystretch}{0.8}
\begin{table}[tb]
\caption{Latency and power for the parameters (\rxc, number of photonic units, $f$, $t_{DAC}$) that minimize EDAP for training and inference under 20 W power and 500 $mm^2$ area.} 
\vspace{-0.1in}
\label{tab:best_params}
\begin{tabular}{l|l|lll|lll}
\hline
\multirow{3}{*}{\textbf{\begin{tabular}[c]{@{}l@{}}Enc.\\ type\end{tabular}}} &
  \multirow{3}{*}{\textbf{Datasets}} &
  \multicolumn{3}{c|}{\textbf{Training}} &
  \multicolumn{3}{c}{\textbf{Inference (1 million)}} \\ \cline{3-8} 
 &
   &
  \multicolumn{1}{l|}{\multirow{2}{*}{Params}} &
  \multicolumn{1}{l|}{\multirow{2}{*}{\begin{tabular}[c]{@{}l@{}}Latency\\ (ms)\end{tabular}}} &
  \multirow{2}{*}{\begin{tabular}[c]{@{}l@{}}Power\\ (W)\end{tabular}} &
  \multicolumn{1}{l|}{\multirow{2}{*}{Params}} &
  \multicolumn{1}{l|}{\multirow{2}{*}{\begin{tabular}[c]{@{}l@{}}Latency\\ (ms)\end{tabular}}} &
  \multirow{2}{*}{\begin{tabular}[c]{@{}l@{}}Power\\ (W)\end{tabular}} \\
 &          & \multicolumn{1}{l|}{} & \multicolumn{1}{l|}{}      &       & \multicolumn{1}{l|}{} & \multicolumn{1}{l|}{}       &       \\ \hline
\multirow{5}{*}{\rotatebox[origin=c]{90}{\textbf{Traditional}}} &
  ISOLET &
  \multicolumn{1}{l|}{\multirow{5}{*}{\begin{tabular}[c]{@{}l@{}}128$\times$76, \\ 4 cores, \\ 5 GHz, \\ 1 ns\end{tabular}}} &
  \multicolumn{1}{l|}{0.09} &
  4.83 &
  \multicolumn{1}{l|}{\multirow{5}{*}{\begin{tabular}[c]{@{}l@{}}128$\times$128, \\ 4 cores, \\ 5 GHz, \\ 1 ns\end{tabular}}} &
  \multicolumn{1}{l|}{8.71} &
  10.34 \\
 & UCIHAR   & \multicolumn{1}{l|}{} & \multicolumn{1}{l|}{0.08}  & 4.86  & \multicolumn{1}{l|}{} & \multicolumn{1}{l|}{8.54}   & 10.17 \\
 & FACE     & \multicolumn{1}{l|}{} & \multicolumn{1}{l|}{6.7}   & 4.96  & \multicolumn{1}{l|}{} & \multicolumn{1}{l|}{8.41}   & 10.38 \\
 & PAMAP    & \multicolumn{1}{l|}{} & \multicolumn{1}{l|}{0.98}  & 4.94  & \multicolumn{1}{l|}{} & \multicolumn{1}{l|}{1.8}    & 9.36  \\
 & PECAN    & \multicolumn{1}{l|}{} & \multicolumn{1}{l|}{0.18}  & 4.73  & \multicolumn{1}{l|}{} & \multicolumn{1}{l|}{5.1}    & 10.01 \\ \hline
\multirow{5}{*}{\rotatebox[origin=c]{90}{\textbf{Record}}} &
  ISOLET &
  \multicolumn{1}{l|}{\multirow{5}{*}{\begin{tabular}[c]{@{}l@{}}128$\times$16, \\ 2 cores, \\ 5 GHz\end{tabular}}} &
  \multicolumn{1}{l|}{0.7} &
  17.26 &
  \multicolumn{1}{l|}{\multirow{5}{*}{\begin{tabular}[c]{@{}l@{}}84$\times$52,\\ 1 core,\\ 5 GHz{]}\end{tabular}}} &
  \multicolumn{1}{l|}{122.45} &
  18.41 \\
 & UCIHAR   & \multicolumn{1}{l|}{} & \multicolumn{1}{l|}{0.63}  & 16.94 & \multicolumn{1}{l|}{} & \multicolumn{1}{l|}{110.04} & 18.61 \\
 & FACE     & \multicolumn{1}{l|}{} & \multicolumn{1}{l|}{56.85} & 17.54 & \multicolumn{1}{l|}{} & \multicolumn{1}{l|}{117.94} & 18.81 \\
 & PAMAP    & \multicolumn{1}{l|}{} & \multicolumn{1}{l|}{9.13}  & 16.46 & \multicolumn{1}{l|}{} & \multicolumn{1}{l|}{20.69}  & 13.5  \\
 & PECAN    & \multicolumn{1}{l|}{} & \multicolumn{1}{l|}{1.24}  & 17.03 & \multicolumn{1}{l|}{} & \multicolumn{1}{l|}{59.44}  & 19.14 \\ \hline
\multirow{3}{*}{\rotatebox[origin=c]{90}{\textbf{Graph}}} &
  DD &
  \multicolumn{1}{l|}{\multirow{3}{*}{\begin{tabular}[c]{@{}l@{}}108$\times$8, \\ 4 cores, \\ 5 GHz\end{tabular}}} &
  \multicolumn{1}{l|}{0.07} &
  14.61 &
  \multicolumn{1}{l|}{\multirow{3}{*}{\begin{tabular}[c]{@{}l@{}}96$\times$48, \\ 1 core,\\ 5 GHz\end{tabular}}} &
  \multicolumn{1}{l|}{52.14} &
  19.86 \\
 & ENZYMES  & \multicolumn{1}{l|}{} & \multicolumn{1}{l|}{0.005} & 11.47 & \multicolumn{1}{l|}{} & \multicolumn{1}{l|}{9.85}   & 12.52 \\
 & PROTEINS & \multicolumn{1}{l|}{} & \multicolumn{1}{l|}{0.01}  & 13.97 & \multicolumn{1}{l|}{} & \multicolumn{1}{l|}{9.14}   & 16.09 \\ \hline
\end{tabular}
\vspace{-0.1in}
\end{table}
\vspace{-0.05in}
\subsection{Power Breakdown of \acceleratorname}\label{subsec:pow-breakdown}

\begin{figure}[tb]
    \centering
    \vspace{-0.05in}
    \includegraphics[width=0.45\textwidth]{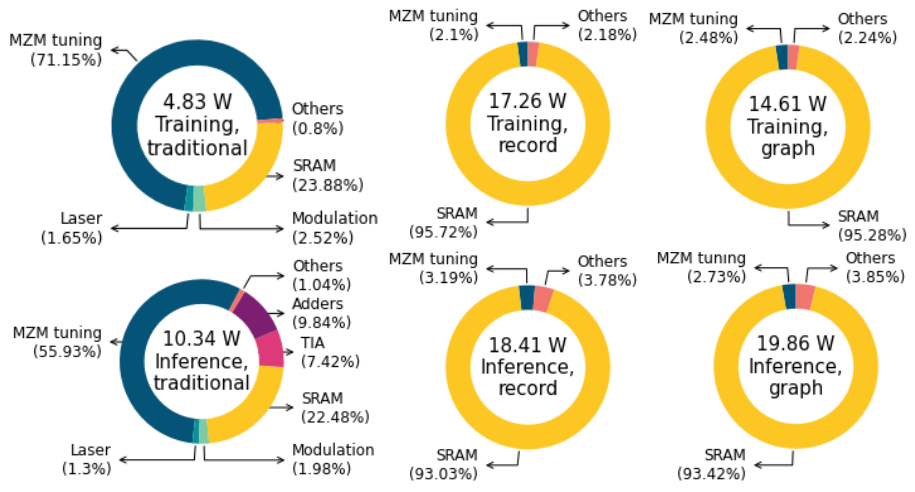}
    \vspace{-0.1in}
    \caption{Power breakdown of \acceleratorname for HDC training and inference for different encoding schemes for ISOLET (\firstEnc and record) and DD (graph). The design parameters are from Table \ref{tab:best_params}.}
    \vspace{-0.25in}
    \label{fig:power_donut}
\end{figure}

In Figure \ref{fig:power_donut}, we show the power breakdown for training and inference with different encoding schemes.
We choose the design parameters listed in Table \ref{tab:best_params} for each plot.
We show the breakdowns for the dataset ISOLET in \firstEnc and record-based encoding and DD for graph encoding as examples.
We observed similar trends for other datasets.

For \firstEnc encoding, we observe that MZM tuning is the dominant (more than 50\%) power-consuming source for both training and inference.
Usually, in electro-photonic DNN accelerators, data converters and lasers are the primary power consumers, while in our design, both the laser power and ADCs contribute very little to the overall power consumption.
This is due to the low precision requirement of HDC (4-bits) that lowers the power consumption exponentially both in the laser source (due to lower SNR requirement) and data converters.
Our short optical path and efficient bundling strategy also help lower the power consumption of the laser source and ADCs.
Therefore, the MZM tuning power becomes the bottleneck in the design.
SRAM data movement is also another major power-consuming source, contributing $\approx$23\% to the total power consumption.
For inference, the power consumption adders and TIAs are higher than in training, as in inference, we use separate adders and TIAs in each row, whereas in training we only use a single adder and TIA for the whole photonic unit.

In record-based encoding, updating tiles every cycle causes a large number of SRAM accesses per cycle resulting in SRAM power consumption being dominant.
The second-most power-consuming source is the MZM tuning circuit, and the other contributions are minimal.
We observed similar trends for graph encoding because it follows the same procedure as record-based encoding.


\subsection{Area Breakdown of \acceleratorname}\label{subsec:area-breakdown}

\begin{figure}[tb]
    \centering
    \includegraphics[width=0.4\textwidth]{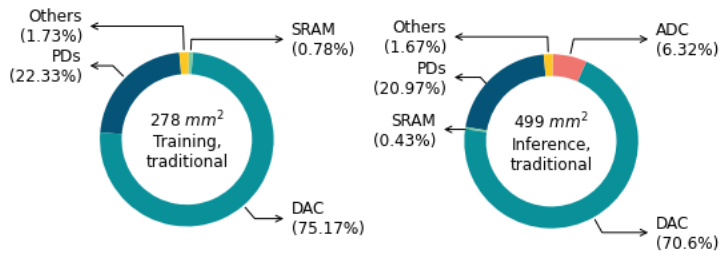}
    \vspace{-0.1in}
    \caption{Area breakdown of HDC training and inference with \acceleratorname for ISOLET. The design parameters are from Table \ref{tab:best_params}.}
    \vspace{-0.2in}
    \label{fig:area_donut}
\end{figure}

In Figure \ref{fig:area_donut}, we show the area breakdown of training and inference with \firstEnc encoding with the architecture parameters (see Table \ref{tab:best_params}) that minimize EDAP for ISOLET.
The breakdown trends are similar in other datasets and other encoding schemes with small variations caused by the size of the SRAM.
For both training and inference, the area is dominated by the DACs.
Each DAC consumes significantly more area than a PD and the DACs account for more than 70\% area of the chip.
For inference, every row of a photonic unit uses an ADC, whereas in training only one ADC is used per photonic unit.
Consequently, ADC contribution is low and it constitutes $\approx$6.3\% of the area in inference and is negligible in training.
The other contributors are MZMs and adders (and ADCs for training), consuming $\approx$2\% of the area.
If one wants to build a single \acceleratorname design that supports both training and inference, then one will need to use the design parameters for inference from Table \ref{tab:best_params} which includes an ADC in every row.


\vspace{-0.05in}
\subsection{Iso-area and Iso-power Analysis}\label{subsec:iso-power-area}
\begin{figure}[t]
    \centering
         \begin{subfigure}[t]{0.21\textwidth}
             \centering
             \includegraphics[width=\textwidth]{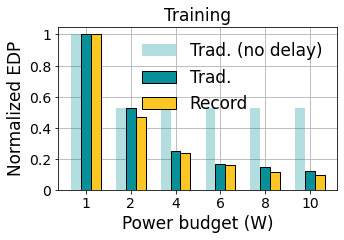}
         \end{subfigure}
         \begin{subfigure}[t]{0.21\textwidth}
             \centering
             \includegraphics[width=\textwidth]{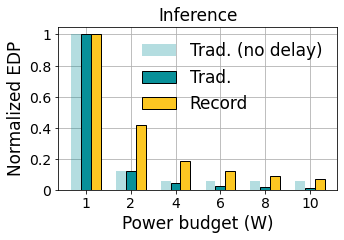}
         \end{subfigure}
    \vspace{-0.1in}
    \caption{Best average EDP (normalized) for different encoding schemes for different power budgets. The area budget is 500 $mm^2$. EDP for \firstEnc encoding with no DAC sharing saturates quickly.}
    \vspace{-0.2in}
    \label{fig:power_budget}
\end{figure}

\begin{figure}[t]
    \centering
         \begin{subfigure}[t]{0.21\textwidth}
             \centering
             \includegraphics[width=\textwidth]{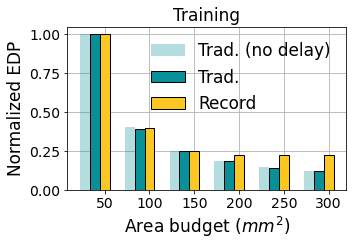}
         \end{subfigure}
         \begin{subfigure}[t]{0.21\textwidth}
             \centering
             \includegraphics[width=\textwidth]{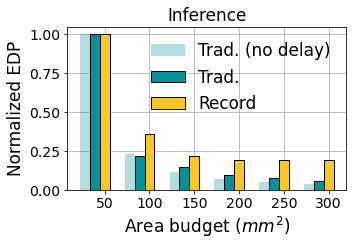}
         \end{subfigure}
    \vspace{-0.1in}
    \caption{Best average EDP (normalized) for different encoding schemes for different area budgets. The power budget is fixed at 20 W. Record-based encoding saturates faster than \firstEnc encoding. 
    }
    \vspace{-0.22in}
    \label{fig:area_budget}
\end{figure}
In this section, for each encoding scheme, we perform an exhaustive search to find the design configuration that minimizes the average EDP across the datasets for different resource (power/area) budgets.
We first plot the normalized average EDP for different power budgets with a fixed 500 $mm^2$ area constraint in Figure \ref{fig:power_budget}.
We observe that the EDP for record-based encoding saturates slowly, meaning that with this encoding and the associated dataflow, the accelerator performance can significantly improve with the increasing power budget.
This is because of the power-hungry nature of the record-based encoding due to frequent SRAM accesses and so the same power budget results in a smaller photonic unit compared to the \firstEnc encoding. 
Therefore, in record-based encoding, allocating a higher power budget allows for larger photonic units and faster execution for both training and inference.
On the other hand, the EDP for \firstEnc encoding with no delays ($t_\text{DAC}=0$) saturates quickly because the large area footprint of DACs goes beyond the 500 mm$^2$ area budget although there is room in the power budget. 
Allowing the extra delay with DAC sharing in \firstEnc encoding helps to lower the DAC area footprint so that the increase in power budget can be used to allocate more compute resources. 

In Figure \ref{fig:area_budget}, we plot the normalized EDP for different area budgets with a fixed 20 W power constraint.
Here, record-based encoding saturates faster because its power consumption uses up the available 20 W power budget even for small-area designs.
However, for \firstEnc encoding, the increased area is utilized without crossing the power constraint.
The extra DAC delay does not make any difference in this case as the area is not the limitation.

The key takeaway from our analysis is that record-based encoding is power-hungry, so having a higher power budget enables higher performance for this encoding scheme.
For \firstEnc encoding, the bottleneck is the area.
Therefore, a larger area is beneficial for achieving better performance.

\vspace{-0.05in}
\subsection{Comparison with SOTA Electro-Photonic DNN Accelerator}\label{subsec:comp-sota-dnn-accel}
\renewcommand{\tabcolsep}{3pt}
\renewcommand{\arraystretch}{0.8}
\begin{table*}[tb]
\centering
\caption{Comparison of SOTA electro-photonic accelerators: ADEPT (AD)~\cite{cansu2021electro}, DEAP-CNN (DC)~\cite{deap-cnn}, and ALBIREO (AL)~\cite{shiflett2021albireo} with \acceleratorname (PH) for HDC training (top) and inference (bottom) using one million samples with \firstEnc encoding. The power consumption of ADCs and DACs of the SOTA accelerators are scaled down to 4 bits for a fair comparison.}
\vspace{-0.05in}
\label{tab:comp_sota_acc_train_inf}
\begin{tabular}{lllllllllllllllll}
\hline
\multicolumn{17}{c}{\textbf{Training}} \\ \hline
\multicolumn{1}{c|}{Datasets} &
  \multicolumn{4}{c|}{Latency (ms)} &
  \multicolumn{4}{c|}{Power (W)} &
  \multicolumn{4}{c|}{EDP (Js)} &
  \multicolumn{4}{c}{Area Efficiency (Sample/s/W/mm2)} \\
\multicolumn{1}{l|}{} &
  AD &
  DC &
  AL &
  \multicolumn{1}{l|}{PH} &
  AD &
  DC &
  AL &
  \multicolumn{1}{l|}{PH} &
  AD &
  DC &
  AL &
  \multicolumn{1}{l|}{PH} &
  AD &
  DC &
  AL &
  PH \\ \hline
\multicolumn{1}{l|}{ISOLET} &
  0.28 &
  2.56 &
  4.36 &
  \multicolumn{1}{l|}{0.09} &
  4.74 &
  71.38 &
  35.82 &
  \multicolumn{1}{l|}{4.83} &
  3.59E-07 &
  4.66E-04 &
  6.81E-04 &
  \multicolumn{1}{l|}{4.00E-08} &
  3.10E+04 &
  8.26E+03 &
  3.43E+02 &
  5.13E+04 \\
\multicolumn{1}{l|}{UCIHAR} &
  0.28 &
  2.54 &
  3.97 &
  \multicolumn{1}{l|}{0.08} &
  4.65 &
  71.38 &
  35.82 &
  \multicolumn{1}{l|}{4.86} &
  3.52E-07 &
  4.62E-04 &
  5.64E-04 &
  \multicolumn{1}{l|}{3.14E-08} &
  3.15E+04 &
  8.26E+03 &
  3.75E+02 &
  5.75E+04 \\
\multicolumn{1}{l|}{FACE} &
  22.30 &
  213.99 &
  365.29 &
  \multicolumn{1}{l|}{6.70} &
  4.49 &
  71.38 &
  35.82 &
  \multicolumn{1}{l|}{4.96} &
  2.23E-03 &
  3.27E+00 &
  4.78E+00 &
  \multicolumn{1}{l|}{2.22E-04} &
  3.38E+04 &
  8.26E+03 &
  3.43E+02 &
  5.69E+04 \\
\multicolumn{1}{l|}{PAMAP} &
  16.73 &
  250.32 &
  55.74 &
  \multicolumn{1}{l|}{0.98} &
  2.56 &
  71.38 &
  35.82 &
  \multicolumn{1}{l|}{4.94} &
  7.17E-04 &
  4.47E+00 &
  1.11E-01 &
  \multicolumn{1}{l|}{4.74E-06} &
  9.24E+04 &
  8.26E+03 &
  2.63E+03 &
  4.57E+05 \\
\multicolumn{1}{l|}{PECAN} &
  0.62 &
  9.13 &
  8.13 &
  \multicolumn{1}{l|}{0.18} &
  4.3 &
  71.38 &
  35.82 &
  \multicolumn{1}{l|}{4.73} &
  1.63E-06 &
  5.95E-03 &
  2.37E-03 &
  \multicolumn{1}{l|}{1.53E-07} &
  5.45E+04 &
  8.26E+03 &
  6.57E+02 &
  9.48E+04 \\ \hline
  \hline
\multicolumn{17}{c}{\textbf{Inference}} \\ \hline
\multicolumn{1}{c|}{Datasets} &
  \multicolumn{4}{c|}{Latency (ms)} &
  \multicolumn{4}{c|}{Power (W)} &
  \multicolumn{4}{c|}{EDP (Js)} &
  \multicolumn{4}{c}{Area Efficiency (Sample/s/W/mm2)} \\
  
\multicolumn{1}{l|}{} &
  AD &
  DC &
  AL &
  \multicolumn{1}{l|}{PH} &
  AD &
  DC &
  AL &
  \multicolumn{1}{l|}{PH} &
  AD &
  DC &
  AL &
  \multicolumn{1}{l|}{PH} &
  AD &
  DC &
  AL &
  PH \\
  \hline
\multicolumn{1}{l|}{ISOLET} &
  42.68 &
  409.60 &
  699.20 &
  \multicolumn{1}{l|}{8.71} &
  5.67 &
  142.76 &
  71.65 &
  \multicolumn{1}{l|}{10.34} &
  1.03E-02 &
  2.40E+01 &
  3.50E+01 &
  \multicolumn{1}{l|}{7.85E-04} &
  1.34E+04 &
  2.07E+03 &
  8.57E+01 &
  2.23E+04 \\
\multicolumn{1}{l|}{UCIHAR} &
  42.68 &
  409.60 &
  638.40 &
  \multicolumn{1}{l|}{8.54} &
  5.59 &
  142.76 &
  71.65 &
  \multicolumn{1}{l|}{10.17} &
  1.02E-02 &
  2.40E+01 &
  2.92E+01 &
  \multicolumn{1}{l|}{7.41E-04} &
  1.36E+04 &
  2.07E+03 &
  9.38E+01 &
  2.32E+04 \\
\multicolumn{1}{l|}{FACE} &
  42.68 &
  409.60 &
  699.20 &
  \multicolumn{1}{l|}{8.41} &
  5.62 &
  142.76 &
  71.65 &
  \multicolumn{1}{l|}{10.38} &
  1.02E-02 &
  2.40E+01 &
  3.50E+01 &
  \multicolumn{1}{l|}{7.35E-04} &
  1.35E+04 &
  2.07E+03 &
  8.57E+01 &
  2.30E+04 \\
\multicolumn{1}{l|}{PAMAP} &
  8.54 &
  409.60 &
  91.20 &
  \multicolumn{1}{l|}{1.80} &
  6.6 &
  142.76 &
  71.65 &
  \multicolumn{1}{l|}{9.36} &
  4.81E-04 &
  2.40E+01 &
  5.96E-01 &
  \multicolumn{1}{l|}{3.03E-05} &
  5.75E+04 &
  2.07E+03 &
  6.57E+02 &
  1.19E+05 \\
\multicolumn{1}{l|}{PECAN} &
  25.61 &
  409.60 &
  364.80 &
  \multicolumn{1}{l|}{5.10} &
  5.73 &
  142.76 &
  71.65 &
  \multicolumn{1}{l|}{10.01} &
  3.76E-03 &
  2.40E+01 &
  9.54E+00 &
  \multicolumn{1}{l|}{2.60E-04} &
  2.21E+04 &
  2.07E+03 &
  1.64E+02 &
  3.94E+04 \\ \hline
\end{tabular}
\vspace{-0.2in}
\end{table*}

Table \ref{tab:comp_sota_acc_train_inf} shows a detailed comparison of the three SOTA electro-photonic DNN accelerators and \acceleratorname for performing HDC training and inference with \firstEnc encoding.
As mentioned in Section \ref{subsec:sota_mapping}, we do not compare the record-based encoding for the SOTA accelerators, as they are all inefficient for this encoding scheme.
We use the architectural configurations prescribed in the corresponding papers of the SOTA accelerators.
For \acceleratorname, we use the parameters mentioned in Table \ref{tab:best_params}.
For calculating area efficiency, we use the area of the active computing elements and the data converters for all accelerators for a fair comparison.
Also, we scale down the power consumption of ADCs and DACs of the SOTA accelerators to 4 bits to match the bit precision of \acceleratorname.

For all datasets, we observe that \acceleratorname achieves better EDP and area efficiency (throughput/power/area) than other accelerators for both training and inference.
ADEPT shows the closest performance to \acceleratorname for training and inference.
In ADEPT, we assume SVD and phase decomposition on the matrix to be programmed in the MZI array are pre-computed. 
While this is possible for single-pass training and inference, ADEPT requires performing SVD after each iteration in the case of iterative training, which can easily become the bottleneck during training. 
In contrast, \acceleratorname can be reprogrammed without any pre-processing of matrices and can be easily used for iterative HDC training.  

We observe that DEAP-CNN is efficient in training if the data fits efficiently into the MRR banks.
However, because the MRRs introduce crosstalk and the tuning circuits are costly, it is challenging to design large MRR banks.
As a result, the total number of computing elements (MRRs) in DEAP-CNN is much smaller than ADEPT (MZIs) and \acceleratorname (PDs), even though DEAP-CNN consumes the highest power.
ALBIREO has the highest latency among all accelerators because it is optimized for CNNs and results in low utilization when running HDC models.
Overall, on average, \acceleratorname achieves three orders of magnitude lower EDP and an order of magnitude better area efficiency for HDC training and two to five orders of magnitude lower EDP and an order of magnitude better area efficiency for HDC inference compared to the SOTA accelerators.

\vspace{-0.05in}
\subsection{Comparison with CiM-based Accelerator}\label{subsec:comparison-cim}
\renewcommand{\tabcolsep}{3pt}
\renewcommand{\arraystretch}{0.9}

\begin{table}[t]
\centering
\caption{Comparison between CiM-based accelerator and \acceleratornametrain (PH) for training and inference with \firstEnc encoding.}
\vspace{-0.05in}
\begin{tabular}{ccccccc}
\hline
\multicolumn{7}{c}{\textbf{Training}}                                 \\ \hline
\multirow{2}{*}{Dataset} &
  \multicolumn{2}{c}{Training time (ms)} &
  \multirow{2}{*}{\begin{tabular}[c]{@{}c@{}}$\times$\\  Speed-up\end{tabular}} &
  \multicolumn{2}{c}{Power (W)} &
  \multirow{2}{*}{\begin{tabular}[c]{@{}c@{}}Power ratio\\ (PH/CiM)\end{tabular}} \\
       & PH       & CiM      &          & PH    & CiM      &          \\
ISOLET & 0.09     & 94.76    & 1.05E+03 & 5.41  & 9.40E-03 & 5.76E+02 \\
UCIHAR & 0.08     & 92.64    & 1.16E+03 & 5.46  & 9.40E-03 & 5.81E+02 \\
FACE   & 6.7      & 6892.52  & 1.03E+03 & 5.58  & 9.40E-03 & 5.94E+02 \\ \hline
\multicolumn{7}{c}{\textbf{Inference}}                                \\ \hline
\multirow{2}{*}{Dataset} &
  \multicolumn{2}{c}{Throughput (inf/s)} &
  \multirow{2}{*}{\begin{tabular}[c]{@{}c@{}}$\times$\\  Speed-up\end{tabular}} &
  \multicolumn{2}{c}{Power (W)} &
  \multirow{2}{*}{\begin{tabular}[c]{@{}c@{}}Power ratio\\ (PH/CiM)\end{tabular}} \\
       & PH       & CiM      &          & PH    & CiM      &          \\ \hline
ISOLET & 1.15E+08 & 7.56E+04 & 1.52E+03 & 11.58 & 1.24E-02 & 9.34E+02 \\
UCIHAR & 1.17E+08 & 7.74E+04 & 1.51E+03 & 11.35 & 1.25E-02 & 9.08E+02 \\
FACE   & 1.19E+08 & 7.59E+04 & 1.57E+03 & 11.62 & 1.25E-02 & 9.30E+02 \\ \hline
\end{tabular}
\vspace{-0.2in}
\label{tab:compare_train_inf_cim}
\end{table}

Table \ref{tab:compare_train_inf_cim} shows the comparison between \acceleratorname and CiM for \firstEnc encoding.
As Neurosim 2.1 does not support bit-precision of less than 5 bits for training, we set the bit-precision to 5 for both CiM and \acceleratorname.
The hyperdimension remains the same, i.e., $D=4096$.
Similar to our previous experiments, we choose the configuration that minimizes the EDAP for \acceleratorname. 

Primarily, due to the fast and energy-efficient multi-bit analog computing capability of photonics, \acceleratorname achieves three orders of magnitude speedup over CiM-based accelerator for both training and inference.
In the CiM approach, multi-bit MVM operations in the ReRAM arrays slow down the HDC operations.
Also, the long reprogramming times of the ReRAM cells make the bundling step time-consuming.
Although the CiM approach is a low-power solution, the speed-up in \acceleratornametrain leads to four orders of magnitude better energy efficiency (EDP) for both training and inference.

\section{Related Work}\label{sec:related_work}
For accelerating HDC, several CPU-based~\cite{imani2017voicehd, imani2019adapthd}, GPU-based~\cite{kang2022openhd, guo2021hyperrec, kim2021cascadehd}, ASIC-based~\cite{khaleghi2022generic, khaleghi2021tiny}, FPGA-based~\cite{salamat2020accelerating, salamat2019f5, imani2019semihd}, and CiM-based~\cite{kazemi2021mimhd,liu2022cosime,zou2022biohd,kazemi2022achieving} solutions have been proposed. 
The FPGA-based techniques generally concentrate on accelerating the encoding step with computation reuse and they have higher performance than the GPU-based implementations~\cite{salamat2020accelerating}.
However, the FPGA resources, such as floating point units, DSP blocks, etc., are substantially more complex than what HDC requires---which calls for a more energy-efficient solution.
Instead, CiM technologies have been extensively explored for HDC applications~\cite{imani2019searchd, barkam2023hdgim, karunaratne2020memory}.
Prior works have demonstrated one or more orders of magnitude better energy efficiency compared to the aforementioned non-CiM designs for HDC.
Nevertheless, inefficient multi-bit operations, high write time, and low endurance issues are still the major challenges in CiM.
    
Over the past few years, researchers have explored the use of photonic computing for accelerating DNNs~\cite{deap-cnn, shiflett2021albireo, sunny2021crosslight, xu202111}. 
These implementations demonstrated multiple orders of magnitude higher compute throughput than their electronic counterparts for MVM operations. 
However, the non-linear operations and the high-precision requirements of DNNs make it challenging to design efficient photonics-based computing systems for DNNs.
From an algorithmic standpoint, HDC is considerably simpler than DNNs, involving primarily linear operations and requiring low precision.
This suggests that photonics is well-suited for HDC than DNNs.
While HDC operations can be mapped onto existing electro-photonic DNN accelerators, they encounter low-utilization, high modulation cost issues, as discussed in detail in Sections \ref{subsec:background_photonics} and \ref{subsec:comp-sota-dnn-accel}.
Therefore, a simplified and tailored design is necessary to efficiently support HDC in electro-photonic systems.

We acknowledge that photonic computing and HDC both are in their early stages of development.
Currently, photonic chips are better suited for cloud and data centers due to limitations in on-chip laser integration and high power consumption, making them particularly attractive for intensive DNNs.
HDC, on the other hand, thrives on its low resource requirements, finding widespread adoption in edge devices.
However, its capabilities are expanding, tackling even heavy data-center workloads as demonstrated in recent research~\cite{zou2022biohd}\cite{baghdadi2021dual}.
Additionally, HDC can be combined with DNNs to increase the noise tolerance and interpretability of DNN models~\cite{imani2017voicehd}.
With HDC getting more popular and requiring more computation, we believe that data centers and cloud systems should consider hardware options such as \acceleratorname for HDC applications.
With developments in laser and photonic device technologies, using photonics on edge devices can also be considered in the near future. 

\section{Conclusion}\label{sec:conclusion}
\vspace{-0.05in}
In this work, we argue that photonics is a suitable hardware platform for HDC.
We propose the first electro-photonic accelerator \acceleratorname for HDC training and inference supporting three popular encoding schemes.
Our results show that for HDC training and inference, \acceleratorname can achieve at least two orders of magnitude lower EDP than SOTA electro-photonic DNN accelerators and four orders of magnitude lower EDP than CiM-based accelerators.

\bibliographystyle{IEEEtranS}
\bibliography{refs}

\end{document}